\documentclass[aps,prl,twocolumn,groupedaddress,showpacs,amsmath]{revtex4}
\usepackage{graphicx}
\usepackage{bm}

\begin{document}

\title{Wave function recombination instability in cold atom interferometers}
\author{James A. Stickney and Alex A. Zozulya}
\email[]{zozulya@wpi.edu}
 \affiliation{Department of Physics, WPI,
100 Institute Road, Worcester, MA 01609}


\begin{abstract}
    Cold atom interferometers use guiding potentials that split
    the wave function of the Bose-Einstein condensate and then
    recombine it.
    We present theoretical analysis of the wave function
    recombination instability that is due to the weak nonlinearity of the condensate.
    It is most pronounced when the accumulated phase difference
    between the arms of the interferometer is close to an odd multiple of $\pi$ and consists in
    exponential amplification of the weak ground state mode by the
    strong first excited mode.
    The instability exists for both trapped-atom and beam interferometers.
\end{abstract}

 \maketitle

Recent experimental demonstrations of miniature cold atom guides
\cite{Denschlag99:guide,Mueller99:guide,Dekker00:guide,Key00:guide},
beamsplitters
\cite{Cassettari00:splitter,Mueller00:splitter,Houde00:splitter,Mueller01:splitter},
and Bose-Einstein condensation (BEC) on a chip
\cite{Hanselnature01:chip} open the possibility of ultra-precise
inertial and rotation measurements via cold atom interferometry.
Cold atom interferometers split the wave function of the
condensate and than recombine it by using guiding potentials that
change from a single well into two separate wells and back. Both
the splitting and recombination are adiabatically slow to avoid
excitation of unwanted modes. The phase of the wave function in
each of the two potential wells evolves independently once the
wells are separated far enough and is sensitive to its local
environment. As a consequence, these two wave functions acquire a
relative accumulated phase difference $\Delta \phi$. When the two
potential wells converge back to a single well, this phase
difference results in an interference that is used to extract
information about the differences between local environments.

In this paper we present theoretical analysis of the wave function
recombination instability that is due to the weak nonlinearity of
the condensate. It is most pronounced when the relative phase
difference $\Delta \phi$ is close to an odd multiple of $\pi$. The
instability consists in exponential amplification of the weak
ground state (symmetric) mode by the strong first excited
(antisymmetric) mode. We calculate the instability growth rate and
present its dependence on the shape of the guiding potential and
the nonlinearity for both trapped-atom and beam interferometers.

Evolution of the condensate in the interferometer will be
described by a one or two-dimensional nonlinear Schr\"{o}dinger
equation (NLSE)
\begin{equation}\label{gp}
    i \frac{\partial}{\partial t}\psi({\bf x},t) =
    \left[-\frac{1}{2}\frac{\partial^{2}}{\partial {\bf x}^{2}} + V({\bf x},t)
        + N |\psi|^{2}\right]\psi({\bf x},t).
\end{equation}
The wave function $\psi$ in Eq.~(\ref{gp}) is normalized to unity,
$V$ is the linear guiding potential, time  is normalized to the
characteristic eigenfrequency $\omega_{0}$ of this potential, and
the spatial coordinates are normalized to the characteristic
length $a_{0} = \left(\hbar/\omega_{0}M\right)^{1/2}$, where $M$
is the atom mass. Finally, $N$ is the normalized nonlinearity
parameter.

In its one-dimensional form, ${\bf x} = x$, Eq.~(\ref{gp})
describes a trapped-atom interferometer
\cite{Hansel01:interf,Reichel01:interf}. The condensate is tightly
confined in two transverse dimensions and is in the lowest
transverse mode of the trap. Equation (\ref{gp}) is the projection
of the Gross-Pitaevskii equation onto this lowest transverse mode.
The nonlinearity parameter $N$ contains the overlap integral
between the three-dimensional wave function of the condensate and
the lowest transverse eigenmode of the trap and is, in general,
time-dependent. In our analysis we assume $N =\mathrm{const}$.
Generalizations to the time-dependent nonlinearity parameter $N$
are straightforward.

In its two-dimensional form, ${\bf x} = (x,y)$, Eq.~(\ref{gp})
describes a beam interferometer
\cite{Hinds01:interf,Andersson02:interf}. Here the condensate
cloud propagates along the z-axis in the guiding potential
$V(x,y,z)$ that confines the condensate in the x-y plane. Equation
(\ref{gp}) is written in a co-propagating frame that is moving
with the condensate. The time dependence of the guiding potential
$V(x,y,t)$ is then parameterized by the longitudinal velocity of
the condensate $v$: $V(x,y,t) = V(x,y,z_{0} + vt)$. The conditions
of applicability of Eq.~(\ref{gp}) are
\begin{equation}
    (Mv/\hbar) = k_{p} \gg (M\omega_{0}/\hbar)^{1/2}
\end{equation}

and
\begin{equation}
    k_{p}l \gg 1,
\end{equation}
where $M$ is the atom mass and $l$ is the characteristic spatial
scale of the guiding potential $V(x,y,z)$ along the $z$ axis. The
first inequality means that the energy associated with the
longitudinal motion of the condensate is much larger than the
characteristic energy $\hbar \omega_{0}$ of transverse eigenmodes
of the interferometer. The second inequality ensures that
undesirable backward reflections of the condensate wave function
from the guiding potential are exponentially small and can be
neglected. Additionally, this inequality allows one to neglect the
$\partial^{2}/\partial z^{2}$ term in Eq.~(\ref{gp}).

We start our analysis from the one-dimensional version of
Eq.~(\ref{gp}) with the simple model guiding potential of the form
\begin{equation}\label{potentialenergy }
    V(x,t) = \left[1 + (\beta(t) - x^{2}/2)^{2}\right]^{1/2},
\end{equation}
where $\beta(t)$ is the control parameter. A zero value of $\beta$
corresponds to a single potential well of the potential $V$.
Positive values of $\beta$ split the potential $V$ into two wells
separated by the distance $d = 2(2\beta)^{1/2}$.

\begin{figure}
\includegraphics{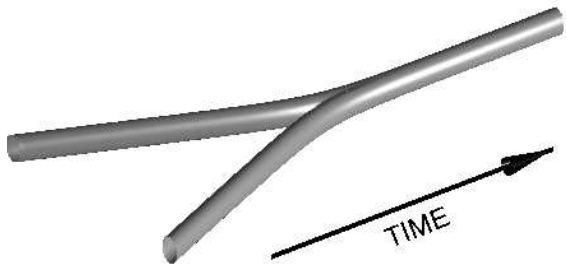}
\caption{\label{fig:splitregionlog} Schematic view of the
recombination region.}
\end{figure}

We assume that the condensate is initially in the lowest weakly
nonlinear mode of the single-well potential $V(\beta=0)$ with the
value of the chemical potential $\mu$ that is about $12\%$ higher
than the eigenenergy of the lowest linear eigenmode $\omega_{1}
\approx 1.3$. This value of $\mu$ corresponds to the nonlinearity
parameter $N = 0.5$ in Eq.~(\ref{gp}). We further assume that the
condensate wave function is split by increasing the control
parameter $\beta$ and acquires a relative phase shift $\Delta \phi
= \pi - 2 \times 10^{-2}$ before the recombination. The
recombination stage is modelled by choosing the control parameter
$\beta(t)$ to be of the form
\begin{equation}\label{controlparameter}
    \beta(t) = A\ln\left[\exp(-t/T) + 1 \right],
\end{equation}
where $A$ and $T$ are constants. Equation (\ref{controlparameter})
describes two separate wells that are linearly converging toward
each other at large negative values of time and merging at large
positive values of $t$. The schematic view of the recombination
region is shown in Fig.~\ref{fig:splitregionlog}.
%
%
\begin{figure}
\includegraphics{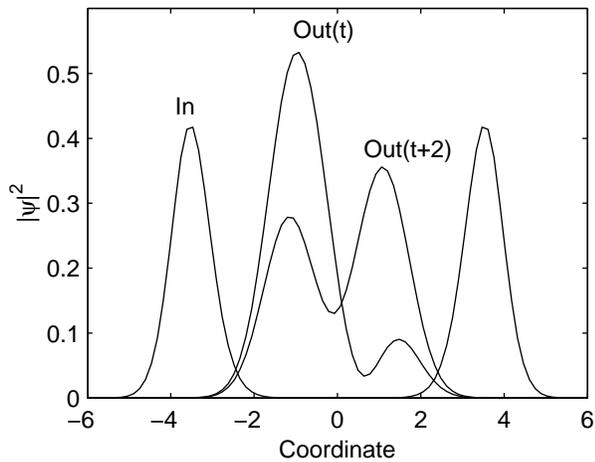}
\caption{\label{fig:intensities} Condensate density before and
after the recombination.}
\end{figure}
%
Figure \ref{fig:intensities} shows the condensate density before
and after the recombination.  The initial wave function is the
combination of the lowest weakly nonlinear eigenmodes of the left
and right potential wells with the relative phase shift $\pi$,
i.e., the first antisymmetric weakly nonlinear eigenmode of the
potential $V$. Because the nonlinearity is weak, this mode
practically coincides with the first antisymmetric linear mode of
the potential $V$. The fractional contributions of the higher
antisymmetric linear modes are about $10^{-4}$. Additionally, the
input wave function contains a small amount of the first linear
symmetric mode of the potential $V$ at the level of $10^{-4}$. The
parameters used for this calculation are $A = 3$ and $T = 90$.

The value of the parameter $T = 90$ ensures that the recombination
is adiabatic and hence the input wave function should map onto the
lowest antisymmetric mode of the single-well potential
$V(\beta=0)$ preserving its odd parity. Figure
\ref{fig:intensities} instead demonstrates that the parity is
broken and the wave function after the recombination is a
superposition of even and odd modes. To quantify this statement we
introduce modal decomposition coefficients $A_{n}(t)$ of the wave
function $\psi(x,t)$ onto linear eigenmodes $\phi_{n}(x,\beta)$ of
the potential $V$ by the relations
\begin{eqnarray}\label{weak:modaldecomposition}
    &&\psi(x,t) = \sum_{n=0}^{\infty}A_{n}(t)
    \phi_{n}(x,\beta(t)), \nonumber \\
    &&A_{n}(t) = \int dx \psi(x,t)\phi_{n}(x,\beta(t)).
\end{eqnarray}
The linear eigenmodes of the potential $V$ $\phi(x,\beta)$ and
their  eigenfrequencies $\omega$ are solutions of the eigenvalue
problem
\begin{equation}\label{eigenequation}
    \omega \phi(x,\beta) = -\frac{1}{2}\frac{d^{2}}{dx^{2}}\phi(x,\beta) +
    V(x,\beta)\phi(x,\beta)
\end{equation}
and parametrically depend on the control parameter $\beta$.
%
\begin{figure}
\includegraphics{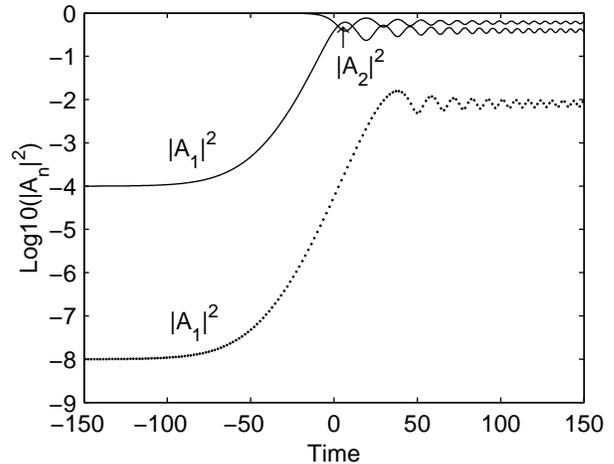}
\caption{\label{fig:modalcoefs} Modal decomposition coefficients
versus time. $A_{1}$ is the ground state mode and $A_{2}$ is the
first excited mode of the guiding potential $V$.}
\end{figure}
%
Figure \ref{fig:modalcoefs} shows modal decomposition coefficients
$|A_{1}|^{2}$ and $|A_{2}|^{2}$ as functions of time on a
logarithmic scale. Solid lines correspond to the results shown in
Fig.~\ref{fig:intensities}. The input wave function shown in
Fig.~\ref{fig:intensities} corresponds to $|A_{2}|^{2} = 1$ and
$|A_{1}|^{2} = \epsilon = 10^{-4}$. Figure \ref{fig:modalcoefs}
clearly demonstrates exponential amplification of the weak mode
$1$ at the expense of the strong mode $2$ until mode $2$ is
depleted. The dotted curve in Fig.~\ref{fig:modalcoefs}
corresponds to a run with the same parameters except the initial
population of the first mode was chosen to be $\epsilon =
10^{-8}$. This curve shows that the instability exists only in
some range of the control parameter $\beta$. If this parameter is
above or below certain values, the system is stable. Indeed, in
the run with $\epsilon = 10^{-8}$ the first mode stabilizes at the
value of $10^{-2}$ after $t \approx 40$ despite the fact that the
mode $2$ (not shown) remains undepleted.

To get an insight into the nature of the instability, we fix the
control parameter $\beta$ at some constant value and replace
Eq.~(\ref{gp}) by a set of coupled equations for the modal
amplitudes $A_{n}(t)$
\begin{equation}\label{a_n}
        i\frac{d}{dt} A_{n} =
        \omega_{n}A_{n} + N
        \sum_{k,l,m}\kappa_{nklm}A_{k}A_{l}^{*}A_{m}.
\end{equation}
Here
\begin{equation}\label{couplingcoefficient}
    \kappa_{nklm} = \int dx \phi_{n}\phi_{k}\phi_{l}\phi_{m}
\end{equation}
are the intermodal overlap integrals.

 Keeping only $A_{1}$ and $A_{2}$ in Eq.~(\ref{a_n})
results in the set of two coupled equations
\begin{eqnarray}\label{2modes}
        &&i \frac{d}{d t} A_{1} =
        \omega_{1}A_{1} +
        N\left(\kappa_{1111}|A_{1}|^{2}A_{1}  \right. \nonumber \\
        &&\left. + 2\kappa_{1122}|A_{2}|^{2}A_{1} +
        \kappa_{1122}A_{2}^{2}A_{1}^{*}\right), \nonumber \\
        &&i \frac{d}{d t} A_{2} =
        \omega_{2}A_{2} +
        N\left(\kappa_{2222}|A_{2}|^{2}A_{2} \right. \nonumber \\
        &&\left. + 2\kappa_{1122}|A_{1}|^{2}A_{2} +
        \kappa_{1122}A_{1}^{2}A_{2}^{*}\right).
\end{eqnarray}
Introducing new variables $y_{1} = 2\mathrm{Re}A_{1}A_{2}^{*}$,
$y_{2} = 2\mathrm{Im}A_{1}A_{2}^{*}$ and $y_{3} = |A_{1}|^{2} -
|A_{2}|^{2}$ transforms Eq.~(\ref{2modes}) to the form
\begin{eqnarray}\label{y}
    &&\frac{d}{dt}y_{1} = \left[- \Delta \omega + N K_{1} +
    N K_{2}y_{3}\right]y_{2}, \nonumber \\
    &&\frac{d}{dt}y_{2} = \left[\Delta \omega - N K_{1} -
     N K_{3}y_{3}\right]y_{1}, \nonumber \\
    &&\frac{d}{dt}y_{3} = -N K_{4} y_{1}y_{2},
\end{eqnarray}
where
\begin{eqnarray}
    &&\Delta \omega = \omega_{2} - \omega_{1}, \nonumber \\
    &&K_{1} = \frac{1}{2}\left(\kappa_{1111} - \kappa_{2222}\right), \nonumber \\
    &&K_{2} = \frac{1}{2}\left(\kappa_{1111} + \kappa_{2222} - 2\kappa_{1122}\right), \nonumber \\
    &&K_{3} = \frac{1}{2}\left(\kappa_{1111} + \kappa_{2222} -
    6\kappa_{1122}\right), \nonumber \\
    &&K_{4} = 2\kappa_{1122}.
\end{eqnarray}
 Equations (\ref{y}) have two
integrals of motion
\begin{eqnarray}\label{integrals}
    &&y_{1}^{2} + y_{2}^{2} + y_{3}^{2} = 1, \nonumber \\
    &&N K_{4} y_{1}^{2} +2 \left[- \Delta \omega + N K_{1}\right]y_{3}
    + N K_{2}y_{3}^{2} = c,
\end{eqnarray}
where $c$ an integration constant, and can be solved in terms of
elliptic functions. In the following we will be interested in the
limit when the amplitude of $A_{2}$ (the first odd eigenmode) is
much larger than that of $A_{1}$ (the first even eigenmode). This
limit corresponds to $y_{1,2} \rightarrow 0$, $y_{3} \rightarrow
-1$ in Eq.(\ref{y}) and yields solutions of the first two
Eq.~(\ref{y}) of the form $y_{1}, y_{2} \propto \exp(\gamma t)$
with $\gamma$ given by the relation
\begin{eqnarray}\label{growthrate}
    &&\gamma^{2} = \left[- \Delta \omega + N(\kappa_{1122}
    -\kappa_{2222})\right] \nonumber \\
    &&\times \left[\Delta \omega + N(\kappa_{2222} -
    3\kappa_{1122})\right].
\end{eqnarray}
The instability corresponds to $\gamma^{2} > 0$.

\begin{figure}
\includegraphics{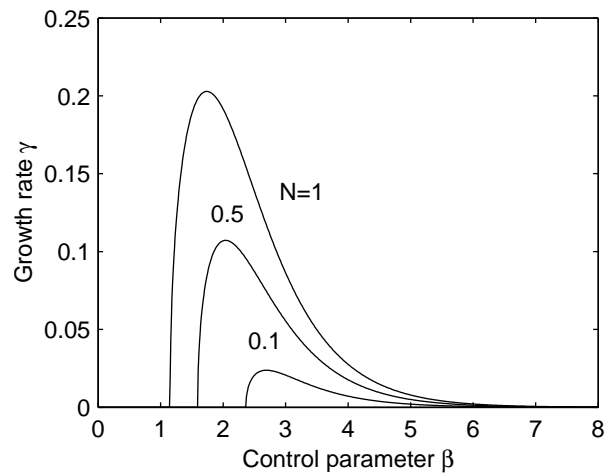}
\caption{\label{fig:growthcurves} Instability growth rate for a
trapped-atom interferometer versus control parameter $\beta$ for
several values of the nonlinearity $N$.}
\end{figure}

Both the eigenfrequencies $\omega$ and the overlap integrals
$\kappa$ in Eq.~(\ref{growthrate}) depend on the shape of the
guiding potential $V$, i.e., on the value of the control parameter
$\beta$.
 Figure \ref{fig:growthcurves} shows the
instability growth rate $\gamma$ as a function of the control
parameter $\beta$ for different values of the nonlinearity $N$.
The curves were obtained by solving the eigenvalue problem
Eq.~(\ref{eigenequation}) and then calculating the growth rate
according to Eq.~(\ref{growthrate}).

%
For each curve plotted in Fig.~\ref{fig:growthcurves} there exists
a cutoff value of the control parameter $\beta$ below which the
system is stable. These results explain saturation of the dotted
curve in Fig.~\ref{fig:modalcoefs}. The maximum growth rate
corresponds to the values of $\beta$ of the order of one. For
large values of $\beta$ the growth rate becomes exponentially
small and scales as $\beta \propto (\Delta \omega)^{1/2}$. The
asymptotic $\beta \gg 1$ corresponds to the frequency difference
$\Delta \omega$ being exponentially small and to all the overlap
integrals $\kappa$ being exponentially close to each other.
Analytical estimates carried out in the framework of a model of
two identical square wells show that formally there exists no
upper boundary on the separation between the wells that makes
$\gamma^{2}$ negative. From the practical point of view though the
instability is quenched once the wells are sufficiently separated
because the growth rate becomes exponentially small.

The preceding analysis has also been carried out with other shapes
of the guiding potential $V$ and gave similar results.
Furthermore, the recombination instability also exists in the case
of beam interferometers with two-dimensional guiding. Results of
the previous modal analysis are straightforwardly carried over to
this case. The instability growth rate is still given by the
Eq.~(\ref{growthrate}). The only difference is that one has to use
the two-dimensional version of the eigenvalue problem
Eq.~(\ref{eigenequation}).

\begin{figure}
\includegraphics{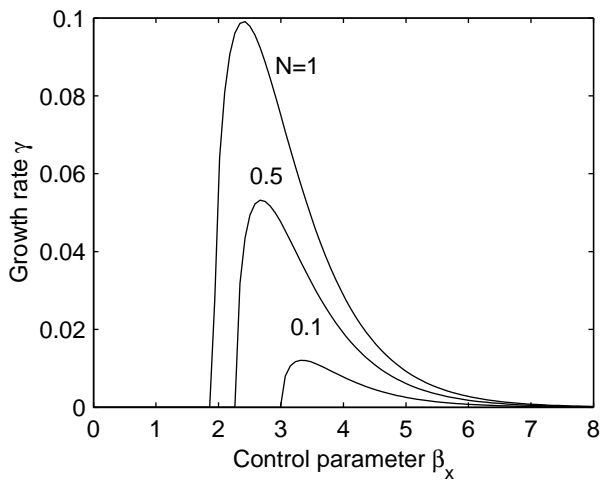}
\caption{\label{fig:growthcurves2d} Instability growth rate for a
beam interferometer versus control parameter $\beta_{x}$ for
several values of the nonlinearity $N$.}
\end{figure}
%
Figure \ref{fig:growthcurves2d} shows the instability growth rate
for the guiding potential
\begin{equation}\label{potential2d}
        V(x,y) = [b_{x}^{2}(x,y) + b_{y}^{2}(x,y) +
        b_{z}^{2}]^{1/2},
\end{equation}
where
\begin{eqnarray}\label{bfield_limit}
    b_{x} &=&\beta_{x}  +\frac{1}{2}(y^{2} - x^{2}), \nonumber \\
    b_{y} &=& xy, \nonumber \\
    b_{z} &=& 1.
\end{eqnarray}
Equation (\ref{potential2d}) describes the guiding potential for
the two-wire cold atom interferometer \cite{Hinds01:interf} in the
limit when the characteristic size of the wave function is much
smaller than the spatial scale of change of the magnetic field.
The $\beta_{x} < 0$ part of the graph is the mirror reflection of
the $\beta_{x} > 0$ part and is not shown. Comparison of
Fig.~\ref{fig:growthcurves} and \ref{fig:growthcurves2d} shows
that the instability exhibits similar qualitative behavior both in
one- and two-dimensional cases.

Results of the preceding analysis demonstrate fascinating dynamics
of weakly nonlinear guided mater waves and offer some insight into
the design of cold atom interferometers. First, the guiding
potential should not be made ''too adiabatic''. Changing the value
of $T$ from $T = 90$ to $T = 30$ ($T = 10$) in
Eq.~(\ref{controlparameter}) reduces the total amplification of
the weak mode from about $10^{6}$ to $10^{2}$ ($10^{1}$) while
still preserving the adiabaticity. Secondly, the control parameter
should be chosen in such a way as to minimize the time the wave
function spends in the region with the largest instability growth
rate.  Finally, the instability analysis results can be used to
estimate practical sensitivity of the interferometer in terms of
errors of determining the phase difference $\Delta \phi$ for a
given level of the nonlinearity.

This work was supported by the U.S. Army Research Office
Multidisciplinary University Research Initiative.



%
\end{document}